\documentstyle[12pt,epsfig]{article} 
 \newcommand{\be}{\begin{eqnarray}}
 \newcommand{\ee}{\end{eqnarray}}
 \newcommand{\beq}{\begin{equation}}
 \newcommand{\eeq}{\end{equation}}
 \newcommand{\ba}{\begin{array}{1}}
 \newcommand{\ea}{\end{array}}
 \newcommand{\bb}{\begin{thebibliography}}
 \newcommand{\eb}{\end{thebibliography}}

 \newcommand{\abstitle}[1]{{\small {\bf #1}}}
 \newcommand{\absauthor}[1]{{\small {\bf #1}}}
 \newcommand{\address}[1]{{\it #1}}
 
\begin{document}
\hfill{LPSC-13-101}
\begin{center}

\abstitle{\large\bf 
Searching for  intrinsic charm in the proton at the LHC
} \\

\vspace{0.4cm}
\absauthor{V.A.~Bednyakov$^1$, M.A.~Demichev$^1$, 
G.I.~Lykasov$^1$, T.~Stavreva$^2$, M.~Stockton$^3$ 
} \\ [0.4cm]

\address{$^1$Joint Institute for Nuclear Research, Dubna 141980, Moscow region, Russia\\ 
         $^2$Laboratoire de Physique Subatomique et de Cosmologie, UJF, 
             CNRS/IN2P3,INPG, 53 avenue des Martyrs, 38026 Grenoble, France\\ 
         $^3$Department of Physics, McGill University, Montreal QC, Canada}

\vspace{0.4cm} 
{\bf Abstract}
\end{center}
     Despite rather long-term theoretical and experimental studies,  
     the hypothesis of the non-zero intrinsic (or valence-like) heavy quark component of the
     proton distribution functions has not yet been confirmed or rejected. 
     The LHC with $pp$-collisions at $\sqrt{s}=$~7--14~TeV 
     will obviously supply extra unique information concerning the above-mentioned 
     component of the proton. 
     To use the LHC potential, first of all, one should select the parton-level (sub)processes 
     (and final-state signatures) that are most sensitive to the intrinsic heavy quark contributions.
     To this end inclusive production of $c(b)$-jets accompanied by photons is considered.
     On the basis of the performed theoretical study 
     it is demonstrated
     that the investigation of the intrinsic heavy quark contributions
     looks very promising at the LHC in processes such as $pp\rightarrow \gamma+  c(b)+X$.

\section{Introduction}
      The Large Hadron Collider (LHC) opens up new and unique kinematical regions with 
      high accuracy for the investigation of the structure of the proton, 
      in particular for the study of the parton distribution functions (PDFs). 
      It is well known that the precise knowledge of the PDFs is essential 
      for the verification of the Standard Model and the search for New Physics. 

      By definition, the PDF $f_a(x,\mu)$ is a function of the proton momentum fraction $x$ 
      carried by parton $a$ (quark $q$ or gluon $g$) at the momentum transfer scale $\mu$. 
      For small values of $\mu$, corresponding to long distance scales less than $1/\mu_0$, 
      the PDF currently cannot be 
 calculated from first principles. 
\cite{LATTICE}. 
      At $\mu>\mu_0$ the $f_a(x,\mu)$  
      can be obtained by means of solving the perturbative QCD evolution equations (DGLAP) 
\cite{DGLAP}.
      At $\mu<\mu_0$ some progress in the calculation of the PDFs
      has been achieved within lattice methods 
\cite{LATTICE}.      
      The unknown (input for the evolution) functions $f_a(x,\mu_0)$ 
      usually can be found empirically from some 
      ``QCD global analysis'' 
\cite{QCD_anal1,QCD_anal2} of a large variety of data typically at $\mu>\mu_0$. 

     In general, almost all $pp$ processes 
     at LHC energies, including Higgs boson production,
     are sensitive to the charm $f_c(x,\mu)$ or bottom $f_b(x,\mu)$ PDFs. 
     Nevertheless, within the global analyses 
     the charm content of the proton at $\mu\sim\mu_c$ and 
     the bottom at $\mu\sim\mu_b$ are both assumed to be negligible.
     Here $\mu_c$ and $\mu_b$ are typical energy scales relevant to the 
     $c$- and $b$-quark QCD excitation in the proton.
     These heavy quark components arise in the proton only perturbatively
     with increases in the $Q^2$-scale 
     through gluon splitting in the DGLAP $Q^2$ evolution 
\cite{DGLAP}. 
     Direct measurement of open charm and open bottom production 
     in deep inelastic processes (DIS) confirms the perturbative 
     origin of heavy quark flavours 
\cite{H1:2005}. 
     However, modern descriptions of these experimental data 
     are not sensitive enough to the 
     above-mentioned perturbative 
     sea heavy quark distributions at relatively 
     large $x$ values ($x>0.1$). 

     Analyzing hadroproduction of the so-called leading hadrons
     Brodsky et al. 
\cite{Brodsky:1980pb, Brodsky:1981} (about thirty years ago)
     have assumed the co-existence of   
     {\it extrinsic} and {\it intrinsic} 
     contributions to the quark-gluon structure of the proton. 
     The {\it extrinsic} (or ordinary) quarks and gluons are generated on 
     a short time scale associated with large-transverse-momentum processes.
     Their distribution functions satisfy the standard QCD evolution equations. 
     The {\it intrinsic} quarks and gluons exist
     over a time scale which is independent of any probe momentum transfer. 
     They can be associated with a bound-state 
{(zero-momentum transfer regime)} hadron dynamics and 
     one believes they have a nonperturbative origin.

       It was shown in 
\cite{Brodsky:1981}
       that the existence of {\it intrinsic} heavy quark pairs 
       $c{\bar c}$, and $b{\bar b}$ within the proton state 
       can be due to the virtue of gluon-exchange and vacuum-polarization graphs. 
       On this basis, 
       within the MIT bag model 
\cite{Golowich:1981}, 
       the probability to find a five-quark component 
       $|uudc{\bar c}\rangle$ bound within the nucleon bag 
       is nonzero and can be about 1--2\%. 

       Initially in 
\cite{Brodsky:1980pb,Brodsky:1981}
       S. Brodsky and coauthors proposed the existence of the 5-quark state $|uudc{\bar c}\rangle$ 
       in the proton. 
       Later some other models were developed. 
       One of them considered a quasi-two-body state 
       ${\bar D}^0(u{\bar c})\, {\bar\Lambda}_c^+(udc)$ in the proton 
\cite{Pumplin:2005yf}. 
       In order to not contradict the DIS HERA data 
       the probability to find the intrinsic charm (IC) in the proton 
       (the weight of the relevant Fock state in the proton)
       was found to be less than 3.5\% 
\cite{Pumplin:2005yf}--\cite{Nadolsky:2008zw}. 
       The probability of finding an intrinsic bottom (IB) state in the proton 
       is suppressed by a factor of $m^2_c/m^2_b\simeq 0.1$ 
\cite{Polyakov:1998rb}, where $m_c\simeq$ 1.3 GeV and $m_b=$ 4.2 GeV are the current 
       masses of the charm and bottom quarks. 
       Therefore, the experimental search for a possible IC signal in $pp$ collisions at 
       the LHC is more promising than the search for the IB contribution.         

       If the distributions of the intrinsic charm or bottom in the 
       proton are hard enough and similar in shape to the valence quark distributions
       (i.e. have valence-like form), then 
one expects 
       the production of charmed (bottom) mesons or charmed 
       (bottom) baryons in the fragmentation region to be similar 
       to the production of pions or nucleons (from the light quarks). 
       However, the yield of this production depends on the probability to find  
       intrinsic charm or bottom in the proton, but this amount looks too small.     
       The PDFs that include the IC contribution in the proton 
       have already been used in perturbative QCD calculations in  
\cite{Pumplin:2005yf}-\cite{Nadolsky:2008zw}.

\begin{figure}[h!] 
\begin{center}
\begin{tabular}{cc}
\epsfig{file=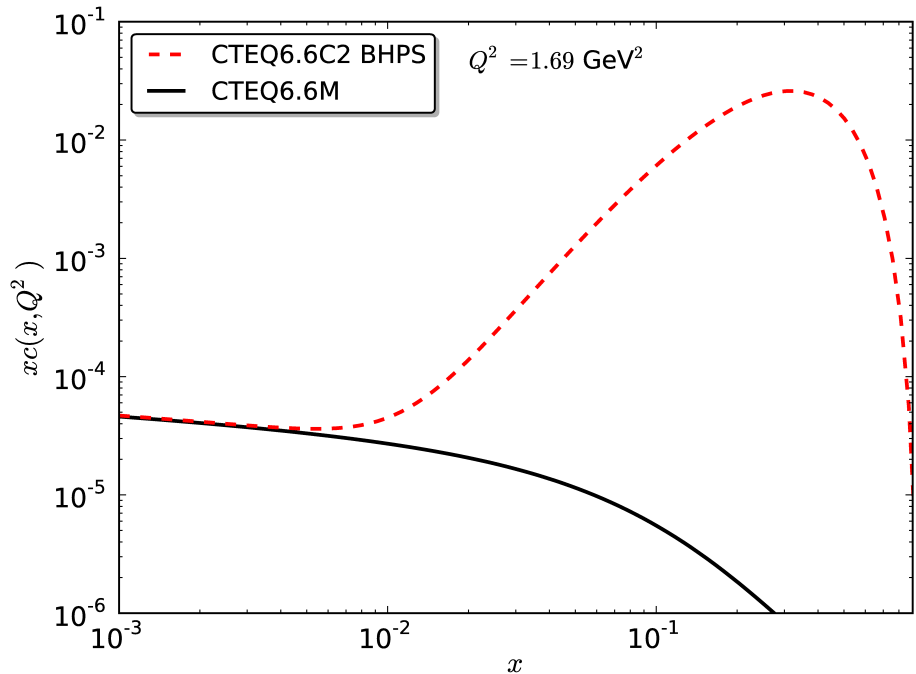,width=0.8\linewidth}
\end{tabular}
\end{center}
\caption{Distributions of the charm quark in the proton. 
  The solid line is the radiatively generated charm density distribution $xc_{\rm rg}(x)$ only, whereas 
  the dashed curve is a full charm quark distribution function, 
  i.e. the sum of the intrinsic charm density $xc_{\rm in}(x)$ 
(see (\ref{def:fcPumpl})) and $xc_{\rm rg}(x)$.} 
\label{Fig_2IC}
\end{figure}

       The probability distribution for the 5-quark state ($|uudc{\bar c}\rangle$) 
       in the light-cone description of the proton was first calculated in 
\cite{Brodsky:1980pb}. 
       The general form for this distribution calculated within the light-cone dynamics
       in the so-called BHPS model 
\cite{Brodsky:1980pb,Brodsky:1981} 
      can be written as 
\cite{Peng_Chang:2012} 
\begin{eqnarray}
P(x_1,..,x_5)=N_5\delta\left(1-\sum_{j=1}^5x_j\right)
\left(m_p^2-\sum_{j=1}^5\frac{m_j^2}{x_j}\right)^{-2}~,
\label{def:B}
\end{eqnarray}
       where $x_j$ is the momentum fraction of the parton, $m_j$ is its mass and $m_p$ is the
       proton mass.  
       Neglecting the light quark ($u,d, s$) masses and the proton
       mass in comparison to the $c$-quark mass and integrating 
(\ref{def:B}) 
       over $dx_1...dx_4$ one can get the probability to find   
       intrinsic charm with momentum fraction $x_5$ in the proton 
\cite{Peng_Chang:2012}: 
\begin{eqnarray}
P(x_5)=\frac{1}{2}{\tilde N}_5x_5^2 \,
\left\{\frac{1}{3}(1-x_5)(1+10x_5+x_5^2)-2x_5(1+x_5)\ln(x_5)\right\}~,
\label{def:fcPumpl}
\end{eqnarray}
     where ${\tilde N}_5=N_5/m^4_{4,5}, m_{4,5}=m_c=m_{\bar c}$, 
     the normalization constant $N_5$ determines some 
     probability $w^{}_{\rm IC}$ to find 
     the Fock state $|uudc{\bar c}\rangle$ in the proton.
 Figure~\ref{Fig_2IC} illustrates the IC contribution in comparison to the 
    conventional sea charm quark distribution in the proton. 

     The solid line in 
Fig.~\ref{Fig_2IC} shows the radiatively generated charm density distribution 
     $xc_{\rm rg}(x)$ (ordinary sea charm) in the proton from CTEQ6.6M
\cite{Nadolsky:2008zw} as a function of $x$ at $Q^2=m_c^2=(1.3)^2$ (GeV$/c)^2$. 
     The dashed curve in
Fig.~\ref{Fig_2IC} is the sum of the intrinsic charm density $xc_{\rm in}(x)$  with the IC probability $w^{}_{\rm IC}= 3.5$\% and $xc_{\rm rg}(x)$ 
      at 
     the same $Q^2=m_c^2$, CTEQ6.6C2 BHPS \cite{Nadolsky:2008zw}.    
      One can see from 
Fig.~\ref{Fig_2IC} 
      that the IC distribution (with $w^{}_{\rm IC} = 3.5$\%) given by 
(\ref{def:fcPumpl}) 
      has a rather visible enhancement at $x\sim $~0.2--0.3 and 
      this distribution is much larger (by a few orders of magnitude) 
      than the sea (ordinary) charm density distribution in the proton. 

     As a rule, the gluons and sea quarks play the key 
     role in hard processes of open charm hadroproduction. 
     Simultaneously, due to the nonperturbative {\it intrinsic} heavy quark components 
     one can expect some excess of these heavy quark PDFs over 
     the ordinary sea quark PDFs at $x>0.1$. 
     Therefore the existence of the intrinsic charm component 
     can lead to some enhancement in the inclusive spectra of open charm hadrons, 
     in particular $D$-mesons, produced at the LHC in $pp$-collisions 
     at large pseudorapidities $\eta$ and large transverse momenta $p_T$ 
\cite{LBPZ:2012}.
    Furthermore, as we know from 
\cite{Brodsky:1980pb}-\cite{Nadolsky:2008zw}
    photons produced in association with heavy quarks $Q(\equiv c,b)$ in the final 
    state of $pp$-collisions provide valuable information about the parton 
    distributions in the proton
\cite{Pumplin:2005yf}-\cite{Thomas:1997}.

\smallskip
    In this paper,
    having in mind these considerations 
    we will 
%
    first discuss where the above-mentioned 
    heavy flavour Fock states in the proton could be searched for 
    at the LHC. 
    Following this 
    we 
    analyze in detail, and give predictions for the LHC 
    semi-inclusive $pp$-production of 
    prompt photons accompanied by  $c$-jets including the {\it intrinsic} 
    charm component in the PDF.  

     For completeness, in these predictions the sea-like charm PDF 
\cite{Pumplin:2007wg} is also considered.  
      As described in 
\cite{Pumplin:2007wg} this is a purely phenomenological scenario, 
     where the intrinsic sea-like charm at the initial scale $Q_0=m_c$ 
     is believed to be proportional to the light sea  PDFs, i.e. 
     $c(x)=\bar c(x) \sim \bar u(x) + \bar d(x)$.  
     This distribution tends to be enhanced at most 
      $x$-values when compared to the ordinary charm distribution (CTEQ6.6M).

\section{The intrinsic charm and beauty}

     According to the model of hard scattering 
\cite{AVEF:1974}--\cite{FF:AKK08}
     the relativistic invariant inclusive spectrum of the hard process 
     $p+p\rightarrow h+X$ can be related to the elastic 
     parton-parton subprocess $i+j\rightarrow i^\prime +j^\prime$,
     where $i,j$ are the partons (quarks and gluons),  
     by the formula
\cite{FF}--\cite{FFF2}:
\begin{eqnarray}
\label{def:rho_c} 
E\frac{d\sigma}{d^3p}= 
\sum_{i,j}\!\int\! d^2k_{iT}\!\int\! d^2k_{jT}\!\int_{x_i^{\min}}^1dx_i\!\int_{x_j^{\min}}^1dx_j
f_i(x_i,k_{iT})f_j(x_j,k_{jT}) 
\frac{d\sigma_{ij}({\hat s},{\hat t})}{d{\hat t}}\frac{D_{i,j}^h(z_h)}{\pi z_h}. 
\label{def:inclsp}
\end{eqnarray}
   Here: $k_{i,j}$ and $k_{i,j}^\prime$ are the four-momenta of the partons $i$ or $j$ 
   before and after the elastic parton-parton scattering, respectively; 
   $k_{iT}, k_{jT}$ are the transverse momenta of the partons $i$ and $j$;  
   $f_{i,j}$ are the PDFs of partons $i$,$j$ inside the proton; 
   $D_{i,j}^h$ is the fragmentation function (FF) of the parton $i$ or $j$ to a hadron $h$; and $z_h$ is the fraction of the final 
   state hadron momentum from the parton momentum.

    When the transverse momenta of the partons are neglected 
    in comparison to the longitudinal momenta, 
    the variables ${\hat s}$, ${\hat t}$, ${\hat u}$ and $z_h$ can be 
    presented in the following form \cite{FF}:\
$
\displaystyle 
{\hat s}=x_i x_j s, \qquad {\hat t}=x_i \frac{t}{z_h}, \qquad
{\hat u}=x_j \frac{u}{z_h}, \qquad  z_h=\frac{x_1}{x_i}+\frac{x_2}{x_j}, \quad
$
     where,
\be
x_1=-\frac{u}{s}=\frac{x_T}{2}\cot({\theta}/{2}), \quad
x_2=-\frac{t}{s}=\frac{x_T}{2}\tan({\theta}/{2}), \quad
x_T=2\sqrt{t u}/s=2p_T/\sqrt{s}.
\ee
      Here as usual, 
      $s=(p_1+p_2)^2$,
      $t=(p_1-p_1^\prime)^2$,
      $u=(p_2-p_1^\prime)^2$, 
     and $p_1$, $p_2$, $p_1^\prime$ are the 4-momenta of the colliding protons 
     and the produced hadron $h$, respectively; 
     $\theta$ is the scattering angle of hadron $h$ in the $pp$ c.m.s.
     The lower limits of the integration in
(\ref{def:rho_c}) are 
\be
x_i^{\min}=\frac{x_T \cot(\frac{\theta}{2})}{2-x_T \tan(\frac{\theta}{2})}, \qquad
x_j^{\min}=\frac{x_i x_T \tan(\frac{\theta}{2})}{2x_i-x_T \cot(\frac{\theta}{2})}.
\label{def:xijmn}
\ee 

     One can see that the Feynman variable $x_F$ of the produced hadron, 
     for example the $D$-meson, can be expressed via  
     the variables $p_T$ and $\eta$, or $\theta$ 
     being the hadron scattering angle in the $pp$ c.m.s: 
\begin{eqnarray}
x_F \equiv \frac{2p_{z}}{\sqrt{s}}
=\frac{2p_T}{\sqrt{s}}\frac{1}{\tan\theta}
=\frac{2p_T}{\sqrt{s}}\sinh(\eta). 
\label{def:xFetapt}
\label{def:xFptteta}
\end{eqnarray} 
        With 
(\ref{def:xFptteta}) the low limit $x_i^{\min}$ in 
(\ref{def:inclsp}) has 
        the following equivalent form:
\be
x_i^{\min}  = \frac{x_R+x_F}{2-(x_R-x_F)}~,    
\label{def:ximin}
\ee
      where $x_R=2p/\sqrt{s}$. 
      One can see from 
(\ref{def:ximin}) that, at least, one of the low limits $x_i^{\min}$ of the integral 
(\ref{def:inclsp}) must be $\geq x_F$. 
      Thus if $x_F\geq 0.1$, then $x_i^{\min}>0.1$, 
      where the ordinary ({\it extrinsic}) charm distribution is completely negligible
      in comparison with the {\it intrinsic} charm distribution. 
      Therefore, at $x_F\geq 0.1$, or equivalently 
      at the charm momentum fraction $x_c> 0.1$ 
      the {\it intrinsic} charm distribution intensifies the charm 
      PDF contribution into charm hadroproduction substantially
(see Fig.~\ref{Fig_2IC}). 
      As a result, 
      the spectrum of the open charm hadroproduction can be increased 
      in a certain region of $p_T$ and $\eta$ (which corresponds 
      to $x_F\geq 0.1$ in accordance to (\ref{def:ximin})). 
      We stress that this excess (or even the very possibility to observe relevant events in this 
      region) is due to the non-zero contribution of the IC component at $x_c > x_F> 0.1$ 
      (where the non-IC component is much smaller).

      This possibility was demonstrated for 
      the $D$-meson production at the LHC in 
\cite{LBPZ:2012}.     
       It was shown that the $p_T$
       spectrum of $D$-mesons is enhanced at pseudorapidities of $3<\eta<5.5$ and 
       10 GeV$/c<p_T<$ 25 GeV/$c$ due to the IC contribution, which was included using the
       CTEQ66c PDF \cite{Nadolsky:2008zw}. For example, due to the IC PDF, with probability about 
       3.5 $\%$, the $p_T$-spectrum increases by a factor of 2 at $\eta=4.5$.
       A similar  effect was predicted in \cite{Kniehl:2012ti}.   

      One expects a similar enhancement in 
      the experimental spectra 
      of the open bottom production 
      due to the (hidden) IB in the proton, which could have  
      a distribution very similar to the one given in
  (\ref{def:fcPumpl}). 
      However, the probability $w_{\rm IB}$ to find
      the Fock state with the IB contribution $|uudb{\bar b}\rangle$  in the proton 
      is about 10 times smaller than the IC probability $w_{\rm IC}$ 
      due to the relation $w_{\rm IB}/w_{\rm IC}\sim m_c^2/m_b^2$ 
\cite{Brodsky:1981,Polyakov:1998rb}.
     
     The IC ``signal'' can be studied not only in the inclusive open (forward) 
     charm hadroproduction at the LHC, but also in some other processes, such as
     production of real prompt photons $\gamma$ or virtual ones $\gamma^*$, or $Z^0$-bosons
     (decaying into dileptons) accompanied by $c$-jets in the kinematics available to the ATLAS and CMS
     experiments.
     The contributions of the heavy quark states in the proton could be 
     investigated also in the $c(b)$-jet 
     production accompanied by the vector bosons $W^\pm,Z^0$. 
     Similar kinematics given by (\ref{def:xFptteta}) and (\ref{def:ximin}) can also be applied to these hard 
     processes.    

     In the next section we analyze in detail the hard process of real photon production in $pp$ collision
     at the LHC accompanied by a $c$-jet including the IC contribution in the proton.      

\section{Prompt photon and $c$-jet production} 

    Recently 
    the investigation of prompt photon and $c(b)$-jet production in
    $p{\bar p}$ collisions at $\sqrt{s}=1.96$~TeV was carried out at the TEVATRON 
\cite{D0:2009}-\cite{Aaltonen:2009wc}.
    In particular,
    it was observed that the ratio of the experimental spectrum of the prompt photons, 
    (accompanied by the $c$-jets) to the relevant theoretical expectation 
    (based on the conventional PDF which ignored the {\it intrinsic} charm) 
    increases with $p_T^\gamma$ up to a factor of about 3 when $p_T^\gamma$ reaches 110 GeV$/c$.
    Furthermore,  
     taking into account 
     the CTEQ66c PDF,
     which includes the IC contribution obtained within the BHPS model \cite{Brodsky:1980pb,Brodsky:1981} 
     one can reduce the difference down to a factor of around 1.5-2  
\cite{Stavreva:2009vi}. 
     For the $\gamma+b$-jets $p{\bar p}$-production no enhancement 
     in the $p_T^\gamma$-spectrum was observed 
     at the beginning of the experiment
\cite{D0:2009,Aaltonen:2009wc}.
     However in 2012 the D\O\ collaboration has confirmed observation of such an enhancement
\cite{Abazov:2012ea}.   

    This intriguing observation stimulates our interest 
    to look for a similar ``IC signal'' in $p p\rightarrow \gamma+c(b)+X$ 
    processes 
    at  the LHC.

\begin{figure}[h!]
\begin{center}
\epsfig{file=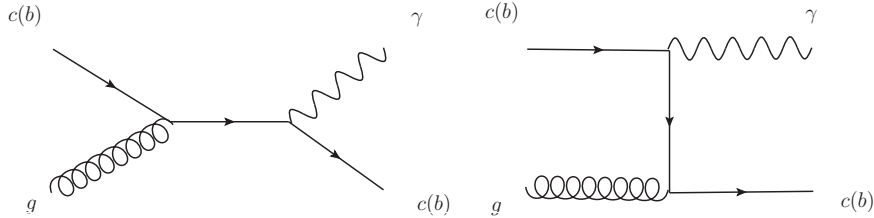,width=0.8\linewidth}
\end{center}
 \caption{The Feynman diagrams for the hard process $c(b) g\rightarrow \gamma c(b)$, the one-quark exchange
  in the s-channel (left) and the same in the t-channel (right).}
\label{Fig_Fd_5}
\end{figure}

    The LO QCD Feynman diagrams for the process $c(b)+g\rightarrow\gamma+c(b)$  
    are presented in 
Fig.~\ref{Fig_Fd_5}.  
    These hard sub-processes give the main contribution to the reaction 
    $pp\rightarrow\gamma + c(b)$-jet$+X$.
   
    The inclusive spectrum of prompt photons 
    produced in $pp$ collisions as a function of their transverse momentum $p_T^\gamma$ 
    is calculated 
    in a similar way to the $D$-meson spectrum 
(\ref{def:rho_c}). 
    The calculatons include the IC contribution through the use of the CTEQ66c PDFs
\cite{Nadolsky:2008zw}. 
    The hard parton-parton cross section (entering 
(\ref{def:rho_c})) 
    for the process $gq\rightarrow \gamma q$ is used in the LO QCD form
\cite{FFF2}: 
\be
\frac{d\sigma_{gq\rightarrow \gamma q}({\hat s},{\hat t})}{d{\hat t}}=
\frac{8\pi}{{\hat s}^2}\alpha_{em}\alpha_s(Q^2)\times\left\{-\frac{e_q^2}{3}
\left(\frac{{\hat u}}{{\hat s}}+\frac{{\hat s}}{{\hat u}}\right)\right\},
\label{def:qgcrsect}
\ee  
    where $\alpha_{\rm em}$ is the electromagnetic coupling constant,
    $\alpha_s(Q^2)$ is the QCD running coupling constant obtained within LO QCD,
    $e_q$ is the quark charge and 
    $Q^2=2{\hat s}{\hat t}{\hat u}/({\hat s}^2+{\hat t}^2+{\hat u}^2)$.
    Within LO QCD, 
    in addition to the main subprocesses illustrated in 
Fig.~\ref{Fig_Fd_5} 
    one considers the 
    subprocesses $gg\rightarrow c{\bar c}$, $q c\rightarrow q c$, $g c\rightarrow g c$ 
    accompanied by the bremstrallung $c({\bar c})\rightarrow c\gamma$,
    the contribution of which is sizable at low $p_T^\gamma$  and can be neglected at
    $p_T^\gamma>$ 60 GeV$/$c, according to \cite{Lipat-Zot:2012}. 
    The diagrams within the NLO QCD are more complicated than 
Fig.~\ref{Fig_Fd_5}. 

    Let us illustrate qualitatively the kinematical regions where 
    the IC component can contribute significantly 
    to the spectrum of prompt photons produced together with a $c$-jet in $pp$ collisions at the LHC. 
    For simplicity we consider only the contribution 
    to the reaction $pp\rightarrow\gamma+c(jet)+X$
    of the diagrams given in 
Fig.~\ref{Fig_Fd_5}. 
    According to (\ref{def:xFetapt}) and (\ref{def:ximin}),
    at certain values of the transverse momentum of the photon, $p_T^\gamma$, 
    and its pseudo-rapidity, $\eta_\gamma$, (or rapidity $y_\gamma$)
    the momentum fraction of $\gamma$ 
    can be $x_{F \gamma}>0.1$, 
    therefore the fraction of the initial $c$-quark must 
    also be above 0.1, where the IC contribution in the proton is enhanced 
(see Fig.~\ref{Fig_2IC}).    
    Therefore, one can expect some non-zero IC signal in the $p_T^\gamma$ 
    spectrum of the reaction $pp\rightarrow\gamma+c+X$ in this
    certain region of $p_T^\gamma$ and $y_\gamma$. 
    In principle, a similar qualitative IC effect can be visible in the production
    of $\gamma^*/Z^0$ decaying into dileptons accomponyed by $c$-jets in $pp$ collisions.

\begin{figure}[ht] 
\begin{center} 
\epsfig{file=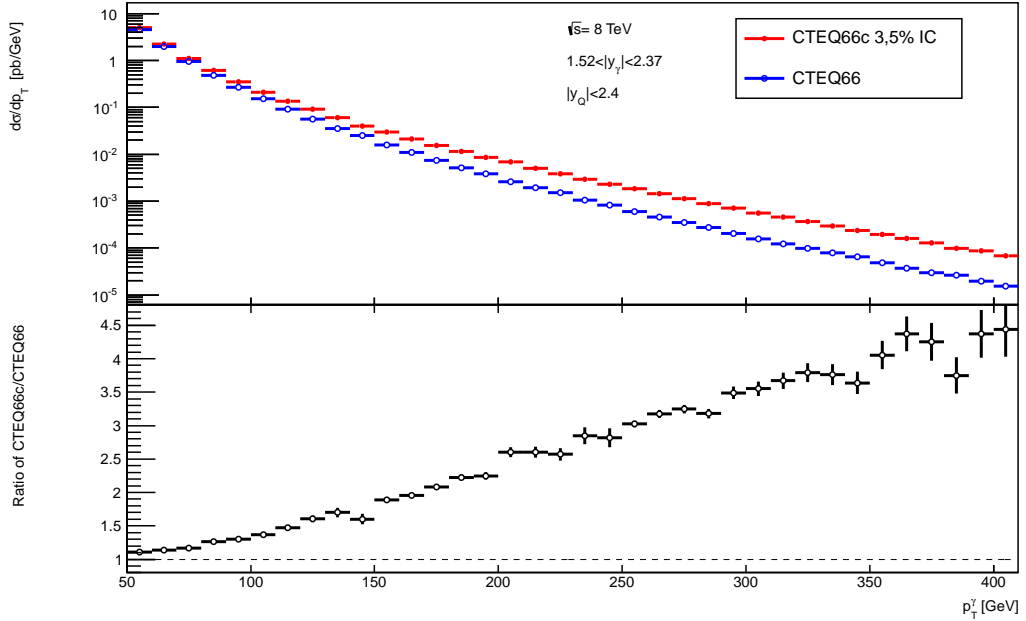,width=1.0\linewidth}
\end{center}
\caption{The PYTHIA8 calculation of the distribution $d\sigma/dp_T^\gamma$  for prompt photons 
  from the reaction $pp\rightarrow\gamma c X$ with transverse momentum $p_T^\gamma$,
  in the interval  1.52$<\mid y_\gamma\mid$<2.37, $\mid y_c\mid<$ 2.4 and at 
  $\sqrt{s}=$ 8 TeV including the hard sub-process $g+c\rightarrow\gamma+c$
  (Fig.~\ref{Fig_Fd_5}). 
  The red solid points (upper points) correspond to the inclusion of
  the IC contribution in the CTEQ66c PDF with IC probability of about 3.5\% 
  \cite{Nadolsky:2008zw};
  the blue open points (lower line) 
  represent the cross-section calculated using the CTEQ66 PDF without the IC 
  contribution. 
}
\label{Fig_diffcrs_1.5}
\end{figure}

In Fig.~\ref{Fig_diffcrs_1.5} the distribution $d \sigma/dp_T^\gamma$ of prompt photons 
       produced in the reaction $pp\rightarrow\gamma+c+X$ at $\sqrt{s}=8$~TeV is presented 
       for the photon rapidity interval $1.52<\mid y_\gamma\mid<2.37$ and for $c$-jet rapidity
       $\mid y_c\mid< 2.4$.
       The calculation was carried out within PYTHIA8 \cite{PYTHIA8} and only the diagrams in Fig.~\ref{Fig_Fd_5} are included.

       The upper line in the top of 
Fig.~\ref{Fig_diffcrs_1.5} is calculated with the use of the CTEQ66c PDF and includes IC,  
       while the lower line uses the CTEQ66 PDF where the charm PDF is radiatively generated only.
       The probability of the IC contribution is about 3.5\%    
\cite{Nadolsky:2008zw}. 
       The ratio of the spectra with IC and without IC as a function of $p_T^\gamma$ is 
       presented in the bottom of Fig.~\ref{Fig_diffcrs_1.5}. 
   
       One can see from 
Fig.~\ref{Fig_diffcrs_1.5} that the inclusion of the IC contribution 
       increases the spectrum by a factor of 4-4.5 at $p_T^\gamma \simeq 400$ GeV$/c$, 
       however the cross section is too small here (about 1 fb). 
       At $p_T^\gamma\simeq$ 150-200 GeV$/c$ the cross section is about 8--30 fb
       if the IC is included and the IC signal reaches 250\% - 300\%. 
       It corresponds to 800--3000 events in the 5 GeV$/$c bin for a luminosity $L = 20$ fb$^{-1}$.

       Naturally the $p_T^\gamma$ distribution in 
Fig.~\ref{Fig_diffcrs_1.5} has the same form as the distribution over
       the transverse momentum of the $c$-quark,  $p_T^c$, 
       when only the hard subprocess $g+c\rightarrow\gamma+c$ in 
Fig.~\ref{Fig_Fd_5} is included.  
 
\begin{figure}[h!]
\begin{center}
\epsfig{file=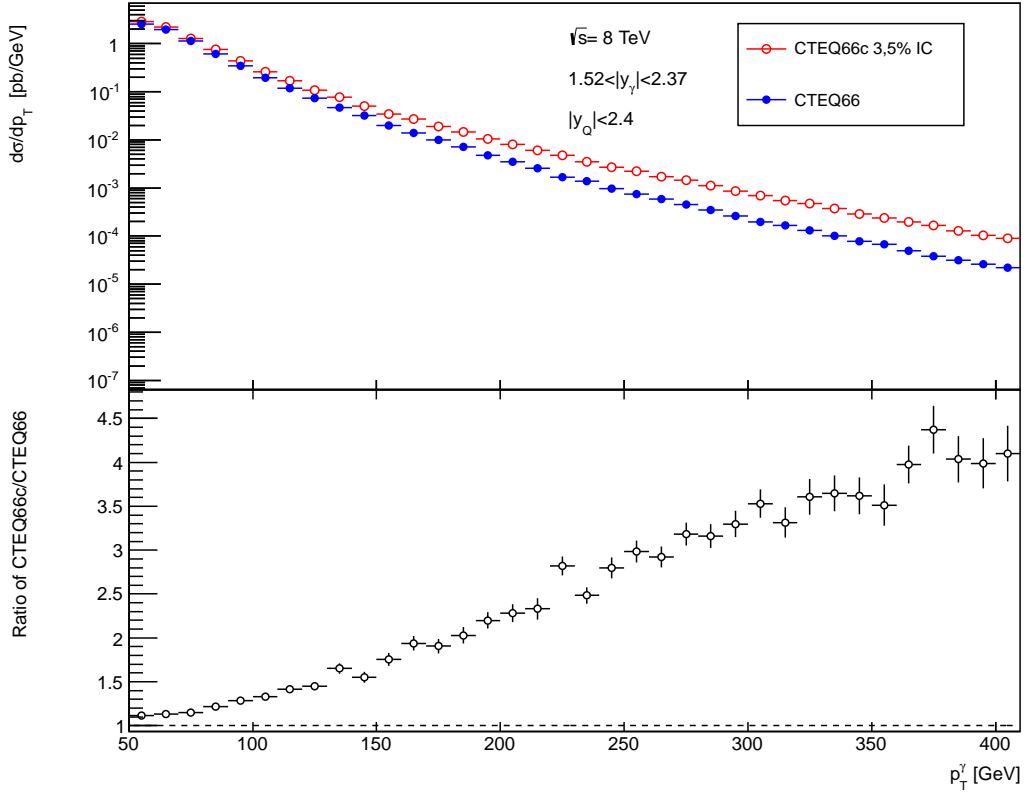,width=1.00\linewidth}
\end{center}
\caption{The distribution $d\sigma/dp_T^\gamma $  
of prompt photons produced in the reaction 
$pp\rightarrow\gamma c X$ over the transverse momentum $p_T^\gamma$
integrated over $dy$ in the interval 1.52$<\mid y_\gamma\mid$<2.37, $\mid y_c\mid<$ 2.4 at 
$\sqrt{s}=$ 8 TeV. The red open points correspond to the inclusion of
the IC contribution in the CTEQ66c PDF with IC probability of about 3.5\% 
\cite{Nadolsky:2008zw};
the blue solid points represent the cross-section calculated using the CTEQ66 PDF without the IC 
contribution. 
The calculation was done within PYTHIA8 using the LO QCD and including the ISR, FSR
and MPI.
}
\label{Fig_diffcrs_2.0}
 \end{figure}

In Fig.~\ref{Fig_diffcrs_2.0} the same distributions as in 
Fig.~\ref{Fig_diffcrs_1.5} are presented 
      including the radiation corrections for the initial (ISR) 
      and final (FSR) states along with the multi-parton 
      interactions (MPI) within PYTHIA8. 

      According to this figure, the IC signal can be about 180\%-250\% 
      at $p_T^\gamma\simeq$ 150-200 GeV$/$c and the cross section
       is about 10--40 fb, which corresponds 
       to about 1000--4000 events in the 5 GeV$/$c bin at $L$=20 fb$^{-1}$.

       Comparing Fig.~\ref{Fig_diffcrs_1.5} and  
Fig.~\ref{Fig_diffcrs_2.0} one can conclude that the inclusion of the
       ISR, FSR and the MPI decreases the cross section at 
       $p_T^\gamma\simeq$ 50-100 GeV$/$c and increases a little bit at
       $p_T^\gamma\simeq>$ 100 GeV$/$c.

\begin{figure}[h!]
\centering{\epsfig{file=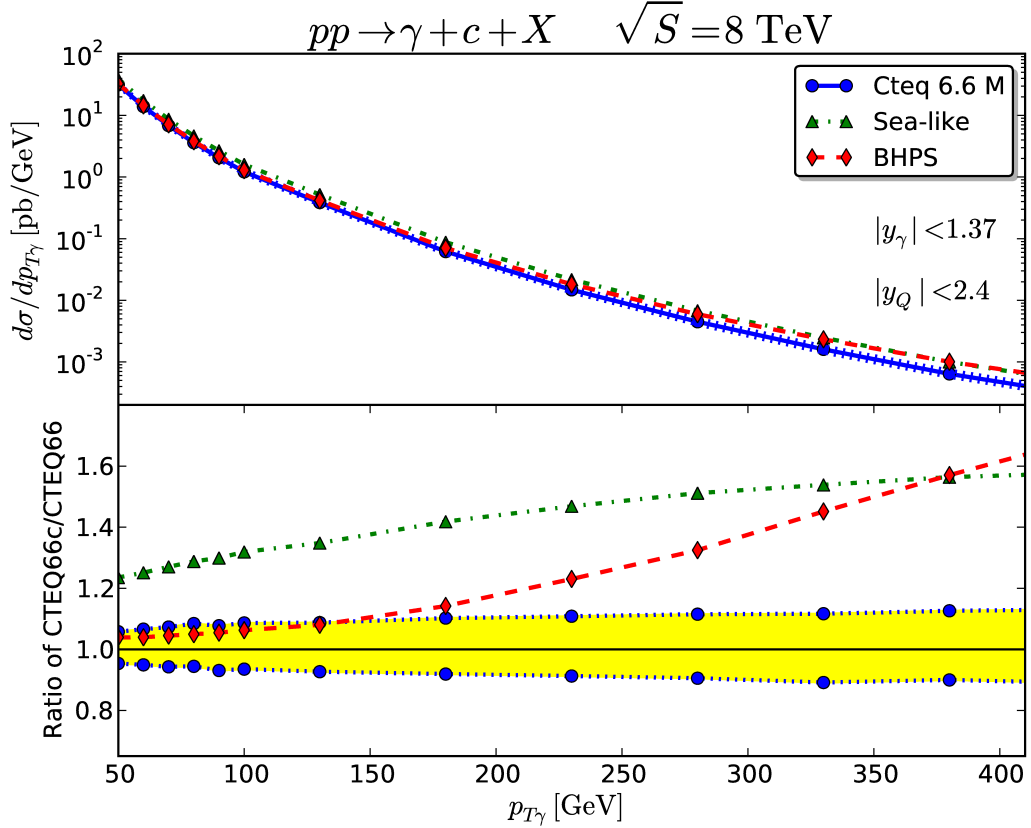,width=0.90\linewidth}}
\caption{
  The $d\sigma/dp_T^{\gamma}$ distribution versus the transverse momentum of the photon for the process $pp\rightarrow\gamma+c+X$
   at $\sqrt{s}=$8 TeV using CTEQ6.6M (solid blue line), BHPS CTEQ6c2 (dashed red line) and sea-like CTEQ6c4 (dash-dotted green line),
   for central photon rapidity $\mid y_\gamma\mid <$1.37 (top).
   The ratio of these spectra with respect to the CTEQ6.6M (solid blue line) distributions (bottom).
   The calculation was done within the NLO QCD approximation.} 
\label{Fig_Tzvt1}
\end{figure}

       In 
Fig.~\ref{Fig_Tzvt1} the differential cross-section $d\sigma/dp_T^\gamma$  calculated at NLO in the massless 
       quark approximation as described in 
\cite{Stavreva:2009vi}  is presented as a function of the transverse momentum of the prompt photon.
       The following cuts are applied: 
       $p_T^\gamma>45$~GeV,  $p_T^c>20$~GeV 
        with the $c$-jet pseudorapidity in the interval $\mid y_c\mid\leq 2.4$ and the photon 
        pseudorapidity in the central region $\mid y_\gamma\mid<1.37$. 

        The solid blue line represents the differential cross-section calculated with the 
        radiatively generated charm PDF (CTEQ66), the dash-dotted green line uses as input 
        the sea-like PDF (CTEQ66c4) 	and the dashed red line the BHPS PDF (CTEQ66c2).
        In the lower half of Fig.~\ref{Fig_Tzvt1} the above distributions normalized to the distribution 
        acquired using the CTEQ66 PDF  and $\mu_r=\mu_f=\mu_F=p_T^\gamma$, are presented.  
        The shaded yellow region, represents the scale dependence.  
        Clearly the difference between the spectrum using the BHPS IC PDF and the one using the radiatively generated PDF 
        increases as $p_T^\gamma$
        increases, however in this central rapidity region at $p_T^\gamma \sim 400$~GeV the BHPS IC and sea-like IC spectra are 
        roughly the same. 
         
\begin{figure}[ht!]
\centering{
\epsfig{file=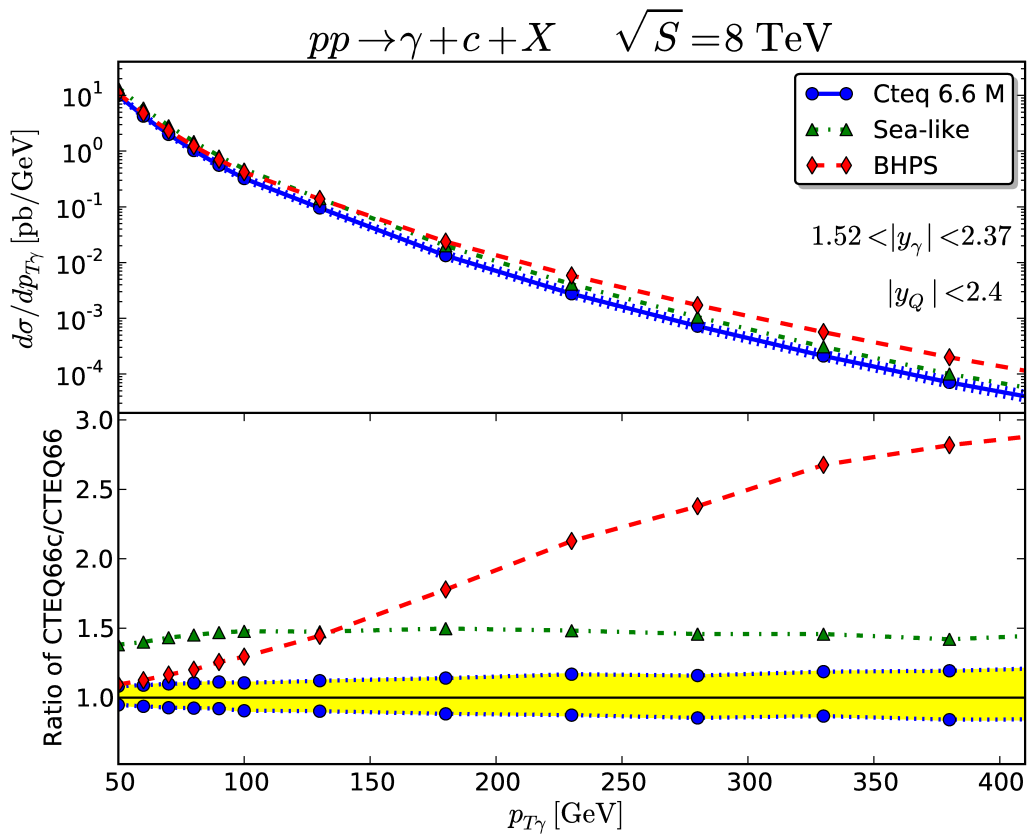,width=0.90\linewidth}}
\caption{
   The $d\sigma/dp_T^{\gamma}$ distribution versus the transverse momentum of the photon for the process $pp\rightarrow\gamma+c+X$
   at $\sqrt{s}=$8 TeV using CTEQ6.6M (solid blue line), BHPS CTEQ6c2 (dashed red line) and sea-like CTEQ6c4 (dash-dotted green line),
   for forward photon rapidity 1.52$<\mid y_\gamma\mid <$2.37 (top).
   The ratio of these spectra with respect to the CTEQ6.6M (solid blue line) distributions (bottom).
   The calculation was done within the NLO QCD approximation. 
} 
\label{Fig_Tzvt2}
\end{figure}

        In Fig.~\ref{Fig_Tzvt2} the same distributions as in Fig.~\ref{Fig_Tzvt1} are shown, however for forward
        photon rapidity $1.52<\mid y_\gamma\mid<2.37$.  In this case larger - $x$ values are probed and therefore 
        we start to observe the difference between the solid and dashed (dash-dotted) lines at smaller $p_T^\gamma$
        values than  in  Fig.~\ref{Fig_Tzvt1}.  The difference when using the BHPS IC PDFs is about 200\% at $p_T^\gamma\sim 200$~GeV
        and increases almost up to 300\% for $p_T^\gamma \sim 400$~GeV.  In this rapidity region the difference between the BHPS and 
        sea-like spectra is clearly visible even as early as  $p_T^\gamma \sim 200$~GeV.  However,  
        while the IC is more 
        accentuated, the cross-section and hence the number of events is less than those for the photon central rapidity in 
        Fig.~\ref{Fig_Tzvt1}.

    Comparing Figs.~\ref{Fig_diffcrs_1.5},\ref{Fig_diffcrs_2.0} to 
    Figs.~\ref{Fig_Tzvt1},\ref{Fig_Tzvt2} one can see that both the LO QCD and NLO QCD 
    cross-section result in approximately the same IC contribution, 
    which increases when the photon transverse momentum grows.
    Nevertheless the values of the spectra calculated within the NLO QCD are larger by a factor 
    of about 1.3 than the ones obtained within the LO QCD at 
    $p_T^\gamma>$ 100 GeV$/$c including the ISR, FSR and the MPI. Note that all the calculations presented in 
    Figs.~(\ref{Fig_diffcrs_1.5}-\ref{Fig_Tzvt2}) were done for isolated photons. 

	Therefore Figs~\ref{Fig_diffcrs_1.5}--\ref{Fig_Tzvt2} show that the 
	IC signal could be visible at the LHC with both the ATLAS and CMS detectors 
	in the process $pp\rightarrow\gamma+c+X$ when $p_T^\gamma\simeq $ 150 GeV$/$c. 
	In this region the IC signal dominates over all non-intrinsic charm background 
	with significance at a level of a factor of 2 (more precisely 170\%).     

\section{Conclusions} 
      In this paper we have shown 
      that the possible existence of the intrinsic heavy quark components in the proton
      can be seen not only in the forward open heavy flavor production in $pp$-collisions
      (as it was believed before) but it can be visible also 
      in the semi-inclusive $pp$-production of prompt photons and $c$-jets
      at rapidities 1.5$<\mid y_\gamma\mid<$ 2.4, $\mid y_c\mid<2.4$ 
      and large transverse momenta of photons and jets.
%
      In the inclusive photon spectrum measured together with a c-jet 
      a rather visible enhancement can appear due to the intrinsic charm  (IC) quark
      contribution. 
      In particular, it was shown that the IC contribution  
      can produce much more events (factor 2 or 3) at $p^\gamma_{T}>$ 150 GeV$/$c 
      and forward $y_\gamma$ 
      in comparison with the the relevant 
      number expected in the absence of the IC. 
      Furthermore the same enhancement is also coherently expected in the  
      transverse momentum, $p_T^c$, distribution of the $c$-jet measured together with 
      the above-mentioned prompt photon in the $p p\rightarrow\gamma + c$-jet$+X$
      process.    
              
      Searching for the signal of {\it intrinsic} charm in such processeses is more pronounced than the search
      for the {\it intrinsic} bottom because the IB probability is, at least, 10 times smaller than the
      IC probability in the proton. 
   
      Our predictions can be verified at the LHC by the ATLAS and CMS Collaborations.
      To this end further consideration of non-charm (light quarks) backgrounds for the discussed 
	processes is mandatory. 

 \vspace{0.4cm}

{\bf Acknowledgments}

We thank S.J. Brodsky and A.A. Glasov for extremely helpful discussions and 
recommendations by the study of this topic. 
The authors are grateful to H. Jung, A.V. Lipatov, V.A.M. Radescu, A. Sarkar and N.P. Zotov for 
very useful discussions and comments.   
This research was also supported by the RFBR grants No. 11-02-01538-a and No. 13-02001060.   
        
\begin{footnotesize}

\end{footnotesize}
\end{document}